\documentclass{article}

\usepackage{arxiv}

\usepackage[utf8]{inputenc} 
\usepackage{adjustbox}
\usepackage{physics}
\usepackage{hyperref}       
\usepackage{url}            
\usepackage{booktabs}       
\usepackage{amsfonts}       
\usepackage{nicefrac}       
\usepackage{microtype}      
\usepackage{lipsum}
\usepackage{graphicx}
\usepackage{float}
\usepackage{chemfig}
\usepackage[version=4,arrows=pgf-filled,
textfontname=sffamily,
mathfontname=mathsf]{mhchem}
\graphicspath{ {./images/} }
\usepackage[T1]{fontenc}    
\title{HI-VQE: Handover Iterative Variational Quantum Eigensolver for Efficient Quantum Chemistry Calculations}
\author{
 Aidan Pellow-Jarman* \And  
 Shane McFarthing \And  Doo Hyung Kang \And  Pilsun Yoo \And 
 Eyuel E. Elala \And  
 Rowan Pellow-Jarman \And  Pratanphorn Nakliang \And  Jaewan Kim \And June-Koo Kevin Rhee
\And Qunova Computing, Inc. \\
{\it Daejeon, Republic of Korea} \\
\\
\texttt{\textsuperscript{*}pellow.a@qunovacomputing.com }
}

\begin{document}
\maketitle
\begin{abstract}

A novel hybrid quantum-classical approach has been developed to efficiently address the multireference quantum chemistry problem. The Handover Iterative Variational Quantum Eigensolver (HI-VQE) is designed to accurately estimate ground-state wavefunctions by leveraging both quantum and classical computing resources. In this framework, noisy intermediate-scale quantum (NISQ) hardwares efficiently explore electron configurations, while classical computers compute the corresponding wavefunction without the effect of noise of NISQ computer, ensuring both accuracy and computational efficiency. By generating compact yet chemically accurate wavefunctions, HI-VQE advances quantum chemistry simulations and facilitates the discovery of novel materials. This approach demonstrates significant potential for overcoming the limitations of classical methods in strongly correlated electronic systems.

\end{abstract}


\section{Introduction}
Quantum computers have the potential to provide vast computational improvements over the best classical methods to solve a variety of problems, particularly the simulation of quantum systems as originally envisioned \cite{Feynman1982}. In recent years, the rapid development of quantum computing hardware has facilitated a renewed push to achieve an exponential speedup over its classical counterparts. One of their most promising applications remains the simulation of quantum systems because of their inherent efficiency in representing the behavior of these systems. In particular, the application of quantum computing to solve the electronic structure problem for molecular systems, where accurate solutions to the non-relativistic time-independent Schrödinger equation are produced \cite{helgaker2013molecular}, could revolutionize the fields of molecular modeling and drug design. 

Though much work has gone into the development of advanced classical computational methods for solving the electronic structure problem \cite{helgaker2013molecular, Ross1952, Gao2024, Cizek1980, Sherrill1999, holmes2016heat, huron1973iterative, liu2016ici, sharma2017semistochastic} over the past several decades, these methods are still limited by the inherent exponential scaling of resources required to simulate molecular systems on classical computers. The Quantum Phase Estimation (QPE) algorithm was proposed as an algorithm to efficiently calculate the ground state energy of a molecular system on a quantum computer \cite{Abrams1999}. However, despite the advancements in quantum hardware in recent years, they still have relatively low coherence times and suffer from large amounts of noise. As a result, quantum algorithms such as QPE, which are highly sensitive to noise and require longer coherence times, will only be applicable in the fault-tolerant era of quantum computing. This has led to the development of hybrid quantum-classical algorithms to leverage the capabilities of the noisy intermediate-scale quantum (NISQ) hardware that's currently available.

The Variational Quantum Eigensolver (VQE) is one such algorithm and was proposed as an alternative to QPE, with the aim of producing accurate ground states for molecular systems on NISQ devices \cite{Peruzzo2014}. It makes use of a hybrid optimization loop that alternates between quantum hardware (the quantum processing unit, QPU) and classical hardware (or the central processing unit, CPU) to generate quantum states which closely approximate the ground state of the system. Parameterized quantum circuits are used to generate trial states, and a classical optimizer updates the circuit parameters to lower the energy of the prepared state until the trial state sufficiently approximates the true ground state for the molecular system. Although VQE was expected to provide a viable near-term alternative to QPE, in the past decade it has become increasingly apparent that it has some fundamental flaws regarding its scalability and trainability \cite{Mcclean2018}. Several algorithms have been developed to address the limitations of the Variational Quantum Eigensolver (VQE). One such approach is the Adaptive Derivative-Assembled Pseudo-Trotter VQE (ADAPT-VQE) \cite{Grimsley2019} along with its numerous variations \cite{Tang2021, Yordanov2021, Halder2024}. These methods primarily focus on improving trainability and reducing circuit depth. However, despite these advancements, they continue to face challenges related to the large number of measurements required to accurately estimate the energy of the generated quantum states.

Recently there has been a shift towards hybrid quantum algorithms that take inspiration from classical Selected Configuration Interaction (SCI) techniques \cite{Sherrill1999, holmes2016heat, liu2016ici, sharma2017semistochastic}. SCI methods were devised as a means to reduce the computational scaling of the Full Configuration Interaction (FCI) method \cite{Ross1952, Gao2024}, a computational approach that provides exact solutions to the electronic structure problem. The FCI method represents the Hamiltonian of a molecular system as a matrix whose elements depend on the interactions between electron configurations of the system, where each distinct configuration represents a possible occupation of the system's orbitals by the electrons. This Hamiltonian matrix is then classically diagonalized, producing its exact eigenvalues and eigenstates, thereby solving the electronic structure problem. Unfortunately, the number of electron configurations in a molecular system scales exponentially with the number of orbitals, such that there is an exponential computational cost required to perform an FCI calculation. As a result, the FCI approach is not feasible for systems with more than a very modest number of orbitals and electrons, even on a supercomputer. 

SCI methods make use of the fact that for many molecular systems, the ground state can be sufficiently well approximated with only a relatively small subset of the system's electron configurations \cite{Szabo1982}, as only these have non-negligible contributions to the ground state energy. The result is that Hamiltonian matrix can be projected into a reduced space of far fewer dimensions when only considering those electron configurations necessary to sufficiently approximate the ground state. However, identifying this subset of electron configurations is non-trivial and various approaches have been developed to construct this so-called subspace \cite{Sherrill1999, holmes2016heat, liu2016ici, sharma2017semistochastic}. Quantum simulation can be a good candidate for efficiently selecting subspaces as investigated in the framework of sampled quantum diagonalizer \cite{Oakridge2025SKQD,bauer2025SQD}.

The Quantum Selected Configuration Interaction (QSCI) algorithm follows the principles of SCI, but uses a quantum computer to generate the subspace of electron configurations \cite{kanno2023quantum}. QSCI prepares a state on a quantum computer and then repeatedly samples the state, obtaining a collection of computational basis states that represent electron configurations for the molecular system. Provided that the state prepared on the QPU device forms a good approximation of the system's ground state, this sampling process can produce the important electron configurations to approximate the ground state. The algorithm then proceeds in the typical SCI fashion by projecting the Hamiltonian into the subspace formed by the electron configurations, and classically diagonalizing the resulting subspace Hamiltonian to obtain the exact ground state within the subspace. An approximate ground state, an example being the result of a VQE calculation, is used in QSCI as the quantum state to improve the likelihood of sampling the important electron configurations. 

Through this process, the trainability issues related to VQE are mitigated because the exact amplitudes of the sampled configurations are classically set, achieving a quantitative accuracy level that is difficult to achieve with VQE \cite{kanno2023quantum}. When sampling the configurations from the quantum state, noise on the QPU device can corrupt some of the samples such that they no longer represent valid electron configurations for the system. To address this, the configuration recovery process was developed to correct corrupted states using a target electron occupation \cite{Robledo2024}.

Recently, concerns were raised that algorithms such as QSCI suffer from poor sampling efficiency, requiring a prohibitive number of quantum measurements to sample the important configurations for larger systems \cite{Reinholdt2025}. Specifically, these issues arise when the sampling of configurations is performed from a state that closely approximates the system's ground state \cite{Reinholdt2025}.

In this paper, we introduce the Handover Iterative Variational Quantum Eigensolver (HI-VQE) algorithm to find the ground state of molecular systems. HI-VQE builds on the principles of the variational quantum eigensolver (VQE), where a quantum simulation circuit controlled by classical parameters produces samples of electron configurations. The VQE algorithm iteratively optimizes these parameters to find a quantum circuit state corresponding to the ground state energy, estimated by the ensemble sum of Pauli word measurements. These measurements sample quantum states from the circuit, directly estimating the molecular system's energy.

\begin{figure}[ht]
\begin{adjustbox}{center}
    \includegraphics[width=\columnwidth]{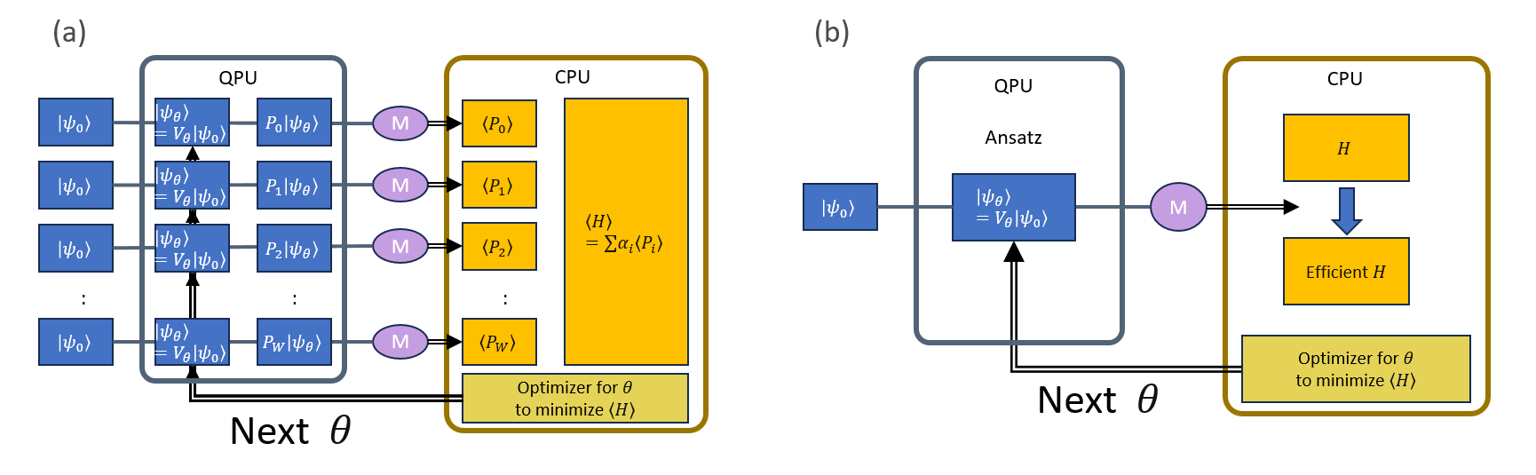}
    \end{adjustbox}
    \caption{Algorithm architecture comparison between (a) conventional VQE algorithm and (b) HI-VQE algorithm. Both algorithms searches the ground state configuration iterative optimizations in the variable space of $\theta$. But the energy estimations are achieved differently by Pauli hamiltonian and hamiltonian eigensolver for VQE and HI-VQE, respectively.}
    \label{fig:1}
\end{figure}

However, VQE has notable drawbacks, as shown in Figure~\ref{fig:1}(a). First, errors in Pauli word measurements directly affect the energy estimate, making VQE inaccurate for modeling molecular systems. Second, VQE requires repeated circuit executions for numerous Pauli word measurements. For instance, a 24-qubit Li$_2$S system requires 15,697 Pauli word measurements to evaluate Hamiltonian energy expectations, necessitating 15,697 circuit runs. Advanced schemes can reduce this number, such as qubit-wise commuting group measurements \cite{Izmaylov2020Pauli}, which reduce the requirement to 4,483 Pauli words. Despite this, the measurement requirement grows rapidly with the number of qubits as $\mathcal{O}(M^4)$ with the size of the system, $M$. In contrast, HI-VQE estimates energy directly from Hamiltonian samples with configurations from a single measurement, as shown in Figure~\ref{fig:1}(b). Consequently, HI-VQE iterations can run thousands of times faster than other VQEs.
Similarly to the QSCI method, HI-VQE is a hybrid quantum-classical algorithm that takes an SCI approach to finding a system's ground state and uses a quantum computer to generate the important electron configurations. However, HI-VQE performs an iterative optimization of the parameters of a quantum circuit to generate a series of quantum states. These states do not necessarily approximate the ground state of the system, instead consisting of distributions of electron configurations that facilitate their efficient sampling, alleviating the concerns of \cite{Reinholdt2025}. This optimization continues, with various processing steps to ensure that only the best configurations are retained, until the ground state energy has converged. The result is the efficient construction of subspaces producing highly accurate ground state wavefunctions.

The remainder of this paper is structured as follows: Section 2 outlines the methods used in this study. Section 3 presents the experimental results and discusses their implications. Finally, Section 4 provides concluding remarks.

\section{Methods}

The Hamiltonian $\hat{H}$ of the molecular system with $M$ spin orbitals and $N$ electrons can be represented as a $2^M\times2^M$ matrix, with the matrix elements.

\begin{equation}
    \hat{H}_{ij} = \langle\psi_i\vert\mathcal{H}_\text{elec}\vert\psi_j\rangle,
    \label{eq:1}
\end{equation}

where the states $\vert\psi_i\rangle$ and $\vert\psi_j\rangle$ belong to an orthonormal basis for the Hilbert space of the Hamiltonian, and $\mathcal{H}_\text{elec}$ is the electronic Hamiltonian,

\begin{equation}
    \mathcal{H}_\text{elec} = -\sum_{i=1}^N\frac{1}{2}\nabla^2_i - \sum_{i=1}^N\sum_{A=1}^M\frac{Z_A}{\vert\mathbf{r}_i-\mathbf{R}_A\vert} + \sum_{i=1}^N\sum_{j>i}^N\frac{1}{\vert\mathbf{r}_i-\mathbf{r}_j\vert}.
\end{equation}

Here, $Z_A$ is the mass of nucleus $A$, $\mathbf{r}_i$ is the position vector of electron $i$ and $\mathbf{R}_A$ is the position vector of nucleus $A$.

The most common orthonormal basis allows each of the basis states to be represented with a corresponding Slater determinant, denoting a specific electron configuration. The matrix elements in Equation \ref{eq:1}, in the basis of Slater determinants can then be calculated efficiently on a classical computer using the Slater-Condon rules \cite{Szabo1982}. 

The FCI method diagonalizes the Hamiltonian matrix $\hat{H}$ to obtain the exact solutions $\ket{\Psi}$ to the non-relativistic time-independent Schrödinger equation, where

\begin{equation}
    \ket{\Psi} = \sum_i\omega_i^*\ket{\psi_i},
    \label{eq:2}
\end{equation}

and each $\omega_i^*$ is the exact amplitude for the basis state $\ket{\psi_i}$ in the target wave function, such as the ground state.

\begin{figure}[ht]
\begin{adjustbox}{center}
    \includegraphics[width=\columnwidth]{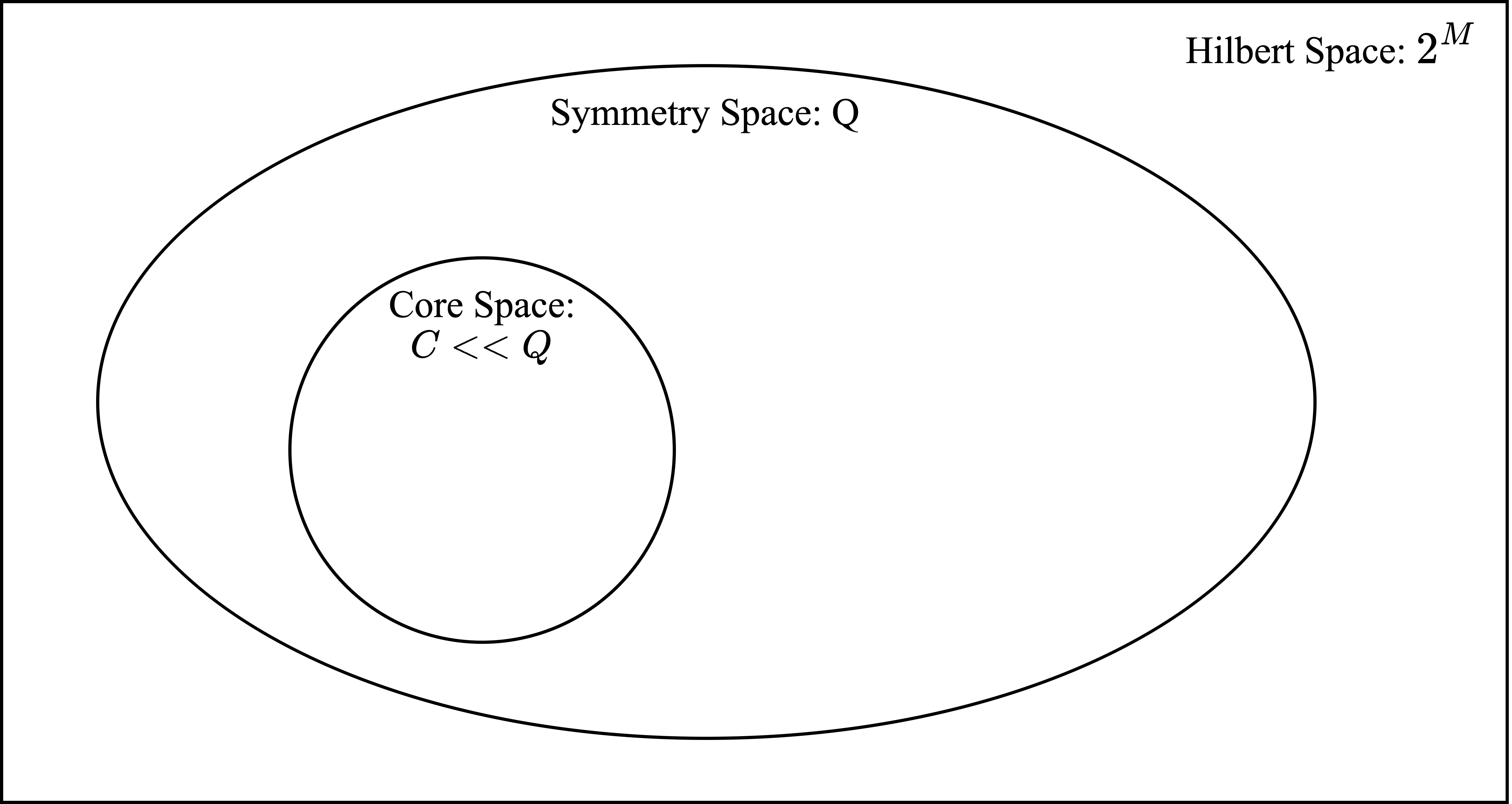}
    \end{adjustbox}
    \caption{The Hilbert space, Symmetry space $Q$, and Core space $C$ for a molecular system of $M$ spin orbitals and $N=N_\alpha+N_\beta$ electrons. The Hilbert space is formed by all possible $2^M$ computational basis states, whereas the Symmetry space $Q$ is selected as the subset of $\binom{M/2}{N_\alpha}\binom{M/2}{N_\beta}$ computational basis states, which forms a set of valid electronic states for the system. The dimension of the Core space $C$ is much smaller than that of $Q$ as it is spanned by the small subset of states from the basis of $Q$ that are sufficient to accurately describe the ground state of the molecular system.}
    \label{fig:2}
\end{figure}

However, for a large number of systems the ground state can be described with sufficient accuracy using only a relatively small subset of the basis states $\ket{\psi_i}$ in the Hilbert space. Some of the possible states in the Slater determinant basis of the Hilbert space do not form valid electronic configurations for the system as they violate its symmetries, such as the number of electrons or the total spin, so we define the Symmetry Space $Q$ to be the space spanned by only those states which preserve the electron number and the total spin. The number of states in $Q$ is defined as

\begin{equation}
    \vert Q \vert = \binom{M/2}{N_\alpha}\binom{M/2}{N_\beta},
    \label{eq:3}
\end{equation}

where $N_\alpha$ and $N_\beta$ are the number of alpha and beta electrons in the system, respectively. We then define the Core space $C$ as the space spanned by the smallest set of states $\ket{\psi_i}$ required to accurately describe the system's ground state. The relationship between these three spaces is depicted in Figure \ref{eq:1}, visualizing the difference in dimension between them.

SCI methods reduce the computational cost of diagonalizing the Hamiltonian matrix by projecting the Hamiltonian into the Core space $C$ using Equation \ref{eq:1}, greatly reducing the dimension of the Hamiltonian matrix $\hat{H}$ while retaining accuracy because of the definition of $C$. In general, it is quite difficult to identify the states that produce a subspace closely resembling $C$, and numerous techniques have been developed for this task \cite{Sherrill1999, holmes2016heat, liu2016ici, sharma2017semistochastic}.

\begin{figure}[ht]
\begin{adjustbox}{center}
    \includegraphics[width=\columnwidth]{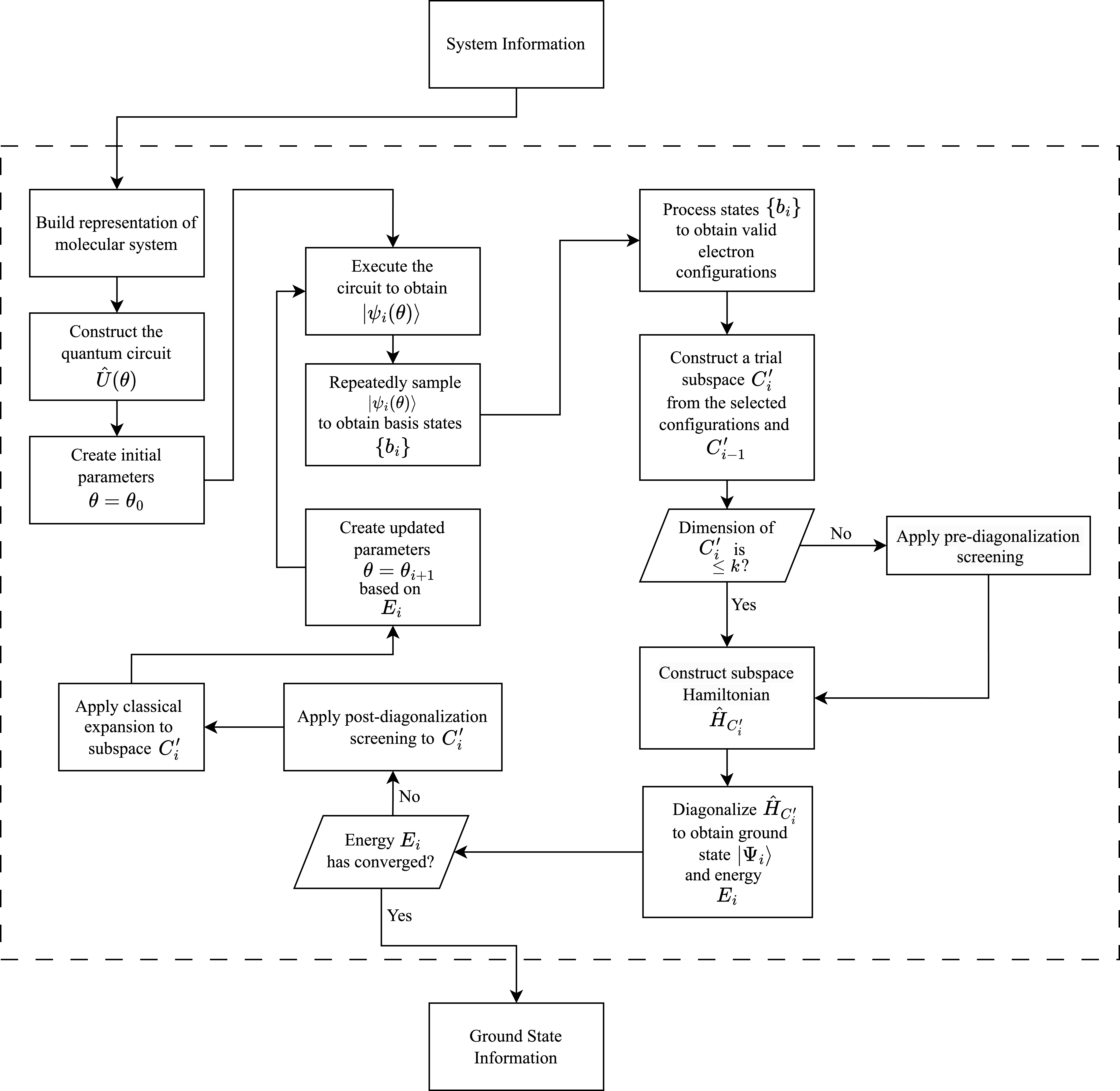}
    \end{adjustbox}
    \caption{An overview of the steps in the HI-VQE algorithm, starting from system information such as the molecule geometry and active space, and ending with the output of the computed ground state information, such as the trial ground state and its energy. The method comprises a hybrid quantum-classical optimization loop where the trial subspace $C'_i$ is iteratively updated to more closely approximate the Core space $C$ of the system. The circuit execution and sampling steps are performed on a quantum computer, and the rest of the steps are performed on classical hardware.}
    \label{fig:3}
\end{figure}

The Handover Iterative Variational Quantum Eigensolver (HI-VQE) method introduced here aims to construct $C$ through an hybrid quantum-classical optimization procedure, where the selection of states in a trial subspace $C'$ is refined. An overview of the steps in HI-VQE is shown in Figure \ref{fig:3}. In this procedure, the quantum device is used to generate the states that form $C'$, and the classical device performs the diagonalization of the subspace matrix resulting from projecting $\hat{H}$ into $C'$. This greatly reduces the required number of quantum measurements compared to traditional VQE methods, as the energy estimation of the trial state is not conducted on the quantum device, instead being classically computed. Additionally, some of the circuit training obstacles in VQE, such as the barren plateau problem, are mitigated in HI-VQE as the quantum state need not quantitatively approximate the true ground state at any iteration.

The exact ground state of the molecular system can be expressed as in Equation \ref{eq:2}, and producing this state on the quantum device with a sufficiently expressible quantum circuit $\hat{U}(\theta)$ requires that the amplitudes $\omega_i$ in the wave function of the resulting quantum state are sufficiently close to, or equal, the values $\omega_i^*$. Producing these optimal amplitudes is the goal of VQE, however trainability issues like barren plateaus makes this very difficult to do at scale \cite{Mcclean2018}. In order to produce the Core space $C$ over a number of iterations, HI-VQE only requires that the amplitudes in the state enable that the corresponding important states should be sampled, quantitative accuracy is not required. Using a classical device to exactly set these amplitudes in the diagonalization of the resulting subspace $C'$ facilitates this reduction in accuracy on the quantum device. Additionally, as the trial subspace $C'$ is cumulatively generated over the optimization process, the quantum state at a given iteration need not approximate the entire ground state wave function, but only a portion of it.

In each iteration of HI-VQE, a quantum state $\ket{\psi_i(\theta)} = \hat{U}(\theta)\ket{\psi_{\text{HF}}}$ is prepared on a quantum device through the application of a parametrized quantum circuit $\hat{U}(\theta)$ to the Hartree-Fock reference state $\ket{\psi_{\text{HF}}}$. We then perform a number of measurements of $\ket{\psi_i(\theta)}$ and obtain the set of computational basis states $\{b_i\}$. These basis states are represented as binary strings and each string represents a specific electron configuration when interpreted as Slater determinants.

Some of the electron configurations sampled from the device can be invalid for the particular system under investigation due to quantum noise, so a filtering step is required to ensure that only configurations which preserve the total spin of the system and the number of electrons are used to construct $C_i'$. The invalid states can either be discarded, or they can be corrected using the recovery process proposed in \cite{Robledo2024}.

A control parameter $k$ is then used to determine whether a pre-diagonalization screening step is required; $k$ specifies a limit for the dimension of the trial subspace $C_i'$, such that configurations will be removed from the subspace if the dimension is too large. The removal of configurations prior to the iteration's final diagonalization of the subspace Hamiltonian requires the estimation of the contribution of each configuration, since the exact amplitudes are not yet available.

In the case that screening is required, HI-VQE projects the Hamiltonian into the subspace $C_i'$ and performs a classical diagonalization with loose termination criteria such that the run time is greatly reduced. This has an appreciable effect on the accuracy of the amplitudes in the resulting ground state wave function, and therefore the energy. However, we find that the returned approximate amplitudes are sufficient for selecting which configurations to discard from the subspace.

The Hamiltonian $\hat{H}$ is then projected into the resulting subspace to obtain the iteration's subspace Hamiltonian $\hat{H}_{C_i'}$, which is subsequently diagonalized on a classical computer. This yields the exact ground state $\ket{\Psi_i}$ within the subspace $C_i'$ and its energy $E_i$. The energy is used to determine if the optimization has converged, and if it has then this information is returned.

Otherwise if it has not converged, the subspace $C_i'$ undergoes an additional screening process using the amplitudes of $\ket{\Psi_i}$. The amplitudes are sorted and any configurations with amplitudes below a set threshold, for example $1\times10^{-6}$, are discarded from the subspace. This aids in keeping the subspace compact, reducing the cost of the classical diagonalization step.

Next, a classical expansion is performed to supplement the quantum generated configurations in the subspace $C_i'$. The purpose of this step is to include any easily obtained configurations that may have been missed during the sampling step. First, the configurations are sorted into descending order based on the magnitude of their amplitudes $\vert\omega_i\vert$ in $\ket{\Psi_i}$. The configuration with the largest $\vert\omega_i\vert$ that has not previously been expanded is then denoted $\ket{\phi_{\text{ref}}}$, and all possible singly and doubly excited configurations $\{\ket{\phi_i}\}$ are generated from $\ket{\phi_{\text{ref}}}$. We then classically calculate $\bra{\phi_{\text{ref}}}\hat{H}\ket{\phi_i}$ for each configuration $\ket{\phi_i}$ using the Slater-Condon rules \cite{Szabo1982}. Only the configurations in $\{\ket{\phi_i}\}$ with the $m$ largest $\bra{\phi_{\text{ref}}}\hat{H}\ket{\phi_i}$ are then added to the subspace $C_i'$. The value for $m$ is specified before HI-VQE begins and dictates the amount of assistance the classical device offers to the quantum subspace generation process.

Finally, a classical optimizer updates the circuit parameters $\theta$ to lower the energy of the computed subspace, and the process is repeated until convergence is achieved or some other predefined termination criteria is met.

During the diagonalization step, the configurations in the subspace $C_i'$ are split into their constituent alpha and beta parts, and the subspace is reconstructed as the tensor product of each alpha configuration with each beta configuration. Additionally, for closed-shell systems it is beneficial to collate the alpha and beta configurations into a single set and take the tensor product of that set with itself to be the subspace, since it is often the case that if a configuration $\alpha\beta$ contributes highly to the ground state, so does $\beta\alpha$. 

Performing this tensor product step can further reduce the sampling cost on the quantum device, but can lead to a rapid increase in the dimension of the trial subspace. As a result, the application of screening protocols with high efficacy at each step is crucial to control the computational cost of the diagonalization step.

In each iteration, two full diagonalizations are performed, one for the constructed trial subspace $C_i'$ and one for the subspace constructed from only the configurations which were sampled in that iteration. Only energy $E_i$ resulting from the second diagonalization is used to update the circuit parameters otherwise the optimizer would be updating the parameters based on an energy calculated from a subspace with configurations from a previous iteration. In this way the circuit is optimized for sampling the most important configurations in the system.

\section{Results and Discussion}

\subsection{Configuration Interaction reference of a few chemical systems}
The number of determinants in a CAS calculation depends on the number of orbitals and electrons, as FCI or CASCI construct wavefunctions based on the available Slater determinants. As shown in Table \ref{tab:1}, the number of determinants increases exponentially with the number of orbitals and electrons. Calculating molecular electronic structures becomes increasingly challenging when the CI matrix requires diagonalization with more than hundreds of millions of states to obtain solutions for the minimum eigenvalue and corresponding eigenvector of a multireference wavefunction.

\begin{table}[ht]

  \caption{Chemical systems and their corresponding properties for Complete Active Space Configuration Interaction Method. The number of determinants is determined with unique electronic configurations with virtual and occupied orbitals with a given number of electrons. The number of qubits are 2 times of the number of orbitals by utilizing the Jordan-Wigner transformation with spin orbitals from spatial orbitals.}
    \centering
    \begin{tabular}{|c|c|c|c|c|c|}
        \hline
        Chemical system & N. of orbitals & N. of electrons & N. of qubits & N. of determinants & Basis function \\ \hline
        \ce{Li2S} & 12 & 12 & 24 & 853,776 & STO-3G \\ \hline
        \ce{NH3}  & 15 & 10 & 30 & 9,018,009 & 6-31G \\ \hline
        \ce{N2}   & 16 & 14 & 32 & 130,873,600 & cc-pVDZ \\ \hline
        \ce{3H2O} & 16 & 16 & 32 & 165,636,900 & 6-31G  \\ \hline
        \ce{N2}   & 20 & 14 & 40 & 6,009,350,400 & cc-pVDZ \\ \hline
        \ce{3H2O} & 22 & 22 & 44 & 497,634,306,624 & 6-31G  \\ \hline
        FePNO     & 22 & 22 & 44 & 497,634,306,624 & 6-31G*  \\ \hline
    \end{tabular}
  
    \label{tab:1}
\end{table}

In the following section, we computed molecular properties such as energy, dipole moments, etc., for these chemical systems to demonstrate the capability of the HI-VQE method. All specified active spaces are created using a HOMO-LUMO selection.

\subsection{HI-VQE for Closed Shell Ground State Calculations}

\subsubsection{Ammonia}

We perform HI-VQE calculations for the chemical systems listed in Table \ref{tab:1} using real hardware available in IBM Quantum. Given that HI-VQE is designed to efficiently search the subspace using quantum hardware, HI-VQE calculations for chemical systems larger than 15 orbitals (or 30 qubits) provide significant advantages. 

\begin{table}[ht]
    \caption{\ce{NH3} properties (Total energy and dipole moment) from RHF, HI-VQE, SHCI and CASCI for 30 qubits or (15o,10e) case.}
    \centering
    \begin{tabular}{|c|c|c|c|c|c|c|}
        \hline
        Method & E$_{total}$ (Ha) & E$_{corr}$ (Ha) & N. of dets & $D_{X}$(Debye) & $D_{Y}$(Debye) & $D_{Z}$(Debye) \\ \hline
        Hartree-Fock  & -56.16003161 & 0.0 & 1 & 0.78861327 & 1.27428558 & -1.83069168 \\ \hline
        HI-VQE ($m=2*10^{3}$) & -56.27829790 & -0.11826629 & 1,936 & 0.81288518 & 1.36670062 & -1.80359989 \\ \hline
        HI-VQE ($m=2*10^{4}$) & -56.28960252 & -0.12957091 & 19,881 & 0.72693956 & 1.19609477 & -1.71483106 \\ \hline
        HI-VQE ($m=2*10^{5}$)  & -56.29215769 & -0.13212608 & 199,809 & 0.77861804 & 1.20181716 & -1.68217164 \\ \hline
        SHCI (eps=$10^{-6}$) & -56.29239971 & -0.13236810 & 1,759,358 & 0.76345000 & 1.23362000 & -1.77228000 \\ \hline    
        CASCI   & -56.29239989 & -0.13236828 & 9,018,009 & 0.76346427 & 1.23364354 & -1.77230786 \\ \hline    
    \end{tabular}
    \label{tab:2}
\end{table}

\begin{figure}[ht]
\begin{adjustbox}{center}
    \includegraphics[width=0.6\textwidth,]{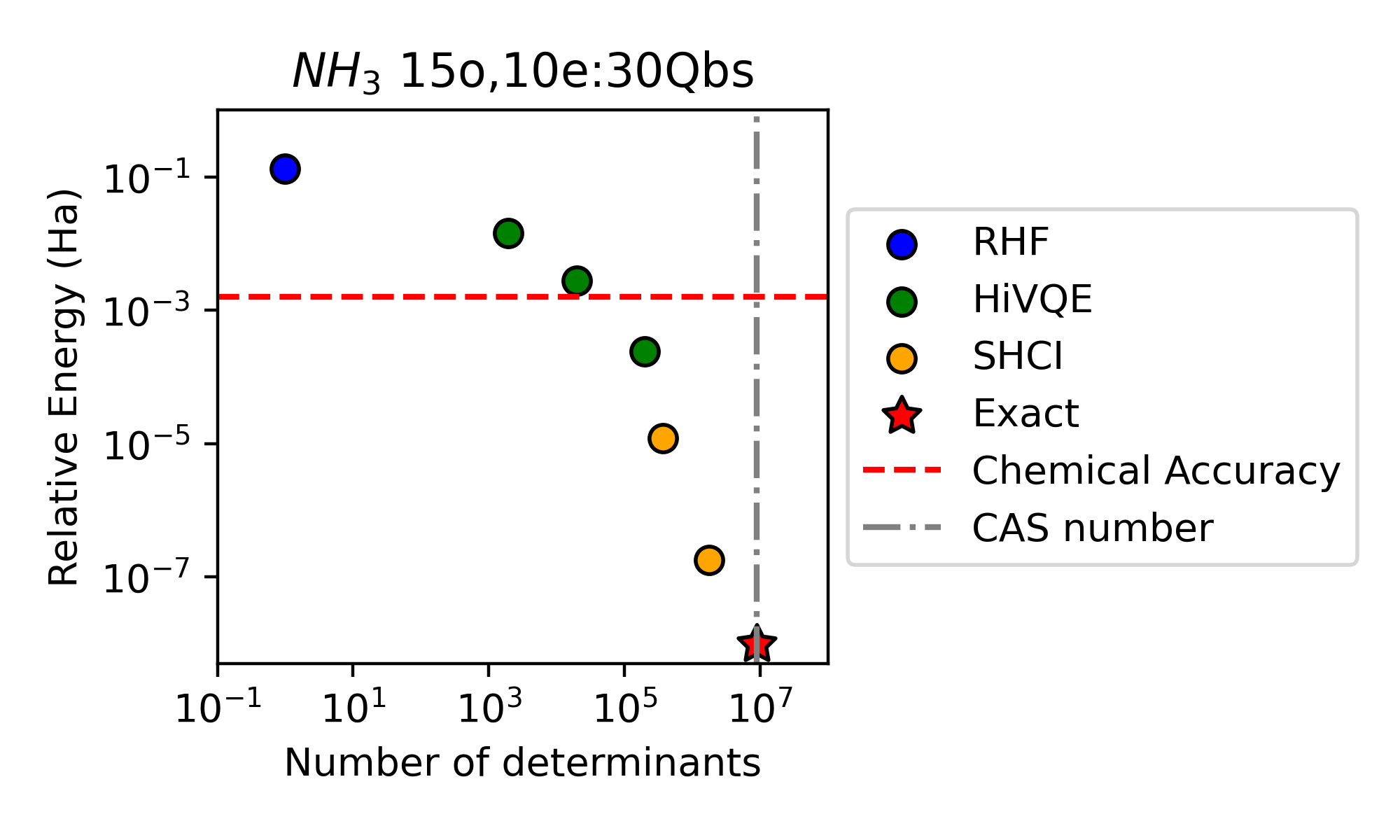}
    \end{adjustbox}
    \caption{The relative energy of RHF, HI-VQE, SHCI to CASCI calculations for \ce{NH3} (15o,10e) system using the 6-31G basis. The x-axis indicates the number of determinants for the CI wavefunction. The CAS number is marked with gray line and Exact (or CASCI) energy is marked at y=$10^{-8}$ Ha to show relative position of the reference within the scale.}
    \label{fig:4}
\end{figure}
\begin{figure}[ht]
\begin{adjustbox}{center}
    \includegraphics[width=0.6\textwidth,]{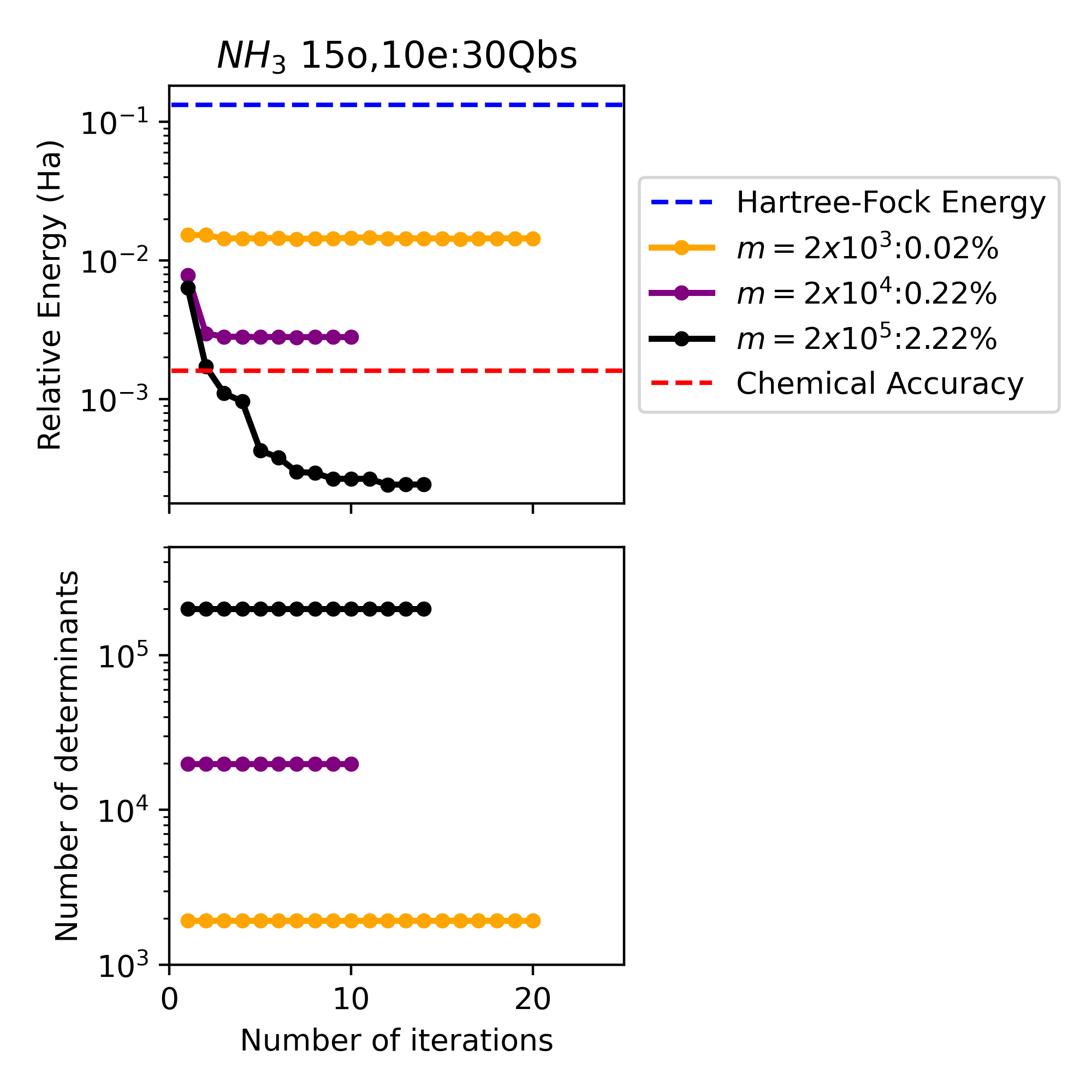}
    \end{adjustbox}
    \caption{The relative energy vs the number of iterations of \ce{NH3} (15o,10e) system using the 6-31G basis with 30 qubit HI-VQE calculations. $m$ is the controlling factor to limit the growth of the subspace matrix. The percentage of the number of states from CAS number is provided together with $m$ parameter. The desired accuracy (<1.6 mHa) can be achieved for \ce{NH3}, 30 qubit system, with only about 2.2\% of the total states.}
    \label{fig:5}
\end{figure}

As a first benchmark calculation, we examined a single \ce{NH3} molecule in gas phase with 30 qubits (15o,10e). It has been reported that clusters of \ce{NH3} exhibits significant static and dynamic electron correlation due to its non-trivial electronic structure, which is missing from Hartree-Fock (HF) or Density Function Theory (DFT). Because the correlation energies can be originated from interactions between occupied and virtual orbitals of multiple atoms, accurate description of the multireference correlation energy contributions of a single molecule would be a good starting point to expand HI-VQE methods for \ce{NH3} cluster studies.
Table \ref{tab:2} shows three individual HI-VQE calculations as well as one of classical SCI approach, Stochastic Heat-bath Configuration Interaction (SHCI)\cite{sharma2017semistochastic} and CASCI as the reference calculation of a single \ce{NH3} molecule for the validation. 

Note that the correlation energy is defined from the difference between Hartree-Fock (HF) energy (E$_{\text{HF}}$) and post-HF energy (E$_{\text{post-HF}}$) as E$_{\text{corr}}$ = E$_{\text{post-HF}}$ - E$_{\text{HF}}$, where post-HF in this study includes SHCI, HI-VQE and CASCI as Exact case. For \ce{NH3} with 30 qubits (15o,10e), the ideal E$_{\text{corr}}$ is calculated as E$_{\text{CASCI}}$ - E$_{\text{HF}}$, yielding -0.13236830 Ha (or -3.60 eV) by incorporating 9,018,008 states in addition to 1 HF state. This highlights the significant computational cost of obtaining correlation energy, emphasizing the need to reduce the number of determinants or states while maintaining accuracy comparable to exact method.
To address this, three HI-VQE calculations were performed, varying $m$ parameter to control the total subspace matrix size (Table \ref{tab:2}). The effect of changing $m$ to the energies is shown in Figure \ref{fig:4}. It demonstrates the convergence of HI-VQE and SHCI towards the exact solution (CASCI). HI-VQE reaches chemical accuracy with fewer determinants compared to SHCI, while RHF exhibits the largest energy deviation from CASCI.

The challenge of configuration interaction lies in the difficulty of accurately estimating the importance of one subset of the subspace over another. To optimize the subspace population sampled from quantum hardware, HI-VQE iteratively examines and selectively filters out less significant states, retaining only those that contribute most. The evolution of the subspace population is evident from the energy changes over iterations, as shown in Figure \ref{fig:5}. Notably, the number of determinants saturates from the beginning in these three cases because the sampled states can exceed the predefined $m$ values. For larger problems exceeding 30 qubits, $m$ can be set to millions to achieve higher accuracy, preventing early saturation during iterations.

It has been observed that HI-VQE with $m=2*10^{5}$ reached the desired precision, which is only 2.22 percent of total space available (Figure \ref{fig:5}). We also note that HI-VQE intends to optimize the quantum ansatz iteratively, limiting the number of samples (or shots) in quantum hardware. Each HI-VQE calculation has several iterations (between 5 to 40 depending on the input parameters). The $m$ parameter, on the other hand, directly affects the precision of the calculations due to the nature of the CI matrix as shown in Figure \ref{fig:5}. With a fixed $m$ value, it still tries to optimize a given size of the CI population by additionally revising the quality of the matrix with lowering the energy from sampled states by quantum circuit. It is clearly shown with the case $m=2*10^5$ where the number of determinants is saturated from the beginning with quantum circuit sampling (the number of shots=1000) and classical expansion to be 100 together. All calculations have been terminated when it reaches a convergence of energy of less than $10^{-5}$ for the last three steps.

\subsubsection{Lithium sulfide}

The precise determination of the chemical reaction energies is essential for progress in fields such as materials science, chemical engineering, and drug discovery. Another system of particular interest is the Li-S bond-breaking process, which plays a key role in the development of advanced battery technologies. In this section, we validate HI-VQE to compute the bond dissociation potential energy surface (PES) of a \ce{Li2S} system by removing a lithium atom. The results can then be compared with reference methods like CASCI and classical approaches such as Hartree-Fock (HF) for a 24-qubit problem.
\begin{figure}[t]
\begin{adjustbox}{center}
    \includegraphics[width=0.5\textwidth,]{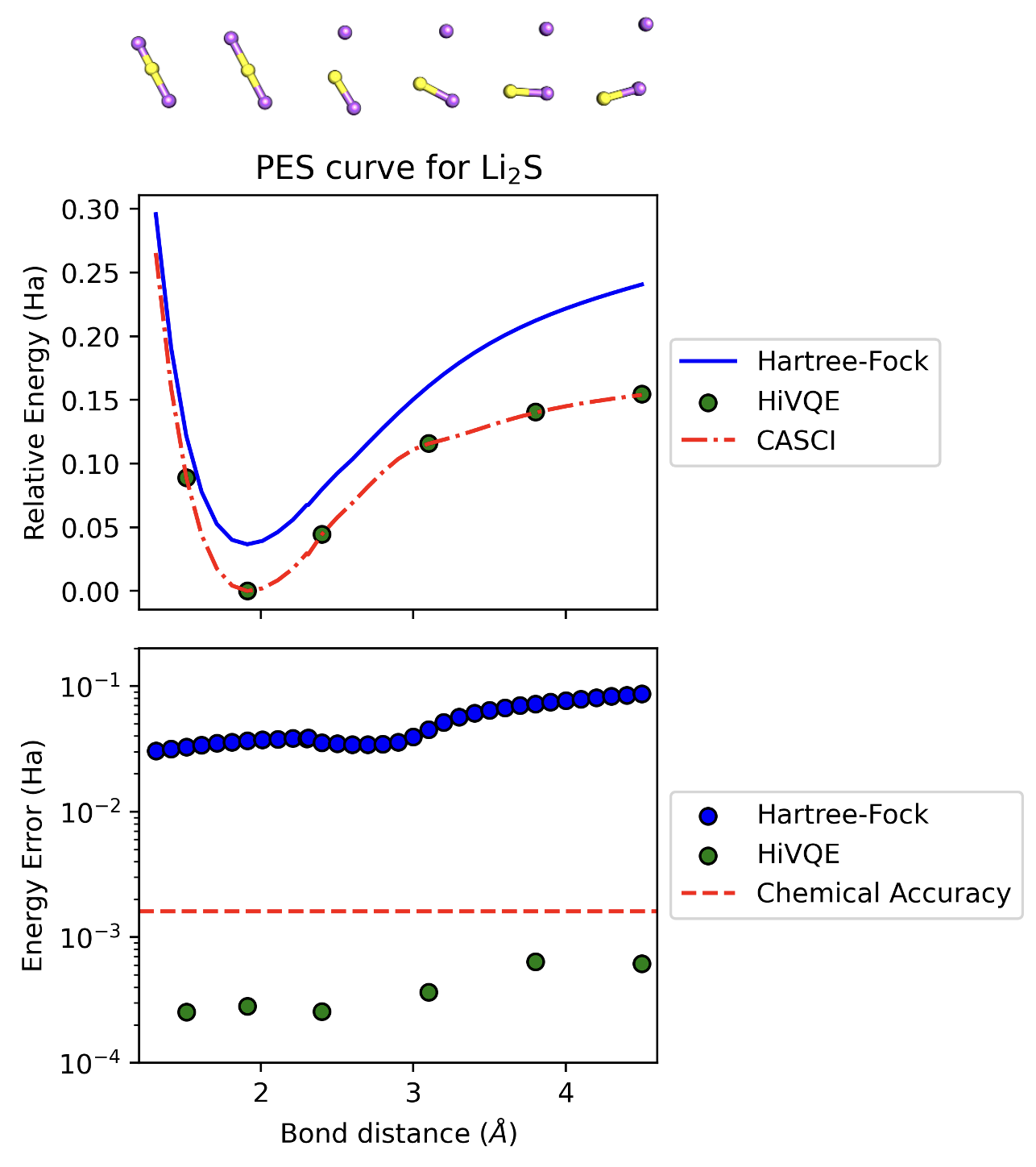}
    \end{adjustbox}
    \caption{The Potential Energy Surface (PES) and energy error analysis for the 
\ce{Li2S} with 24 qubits or (12o,12e). The top panel compares the relative energy computed using Hartree-Fock (blue line), HI-VQE (green circles), and CASCI (red dashed line) as a reference with the corresponding geometries for each energy point at the top. The bottom panel presents the energy error relative to CASCI on a logarithmic scale.}
    \label{fig:6} 
\end{figure}

\begin{figure}[H]
\begin{adjustbox}{center}
    \includegraphics[width=0.5\textwidth,]{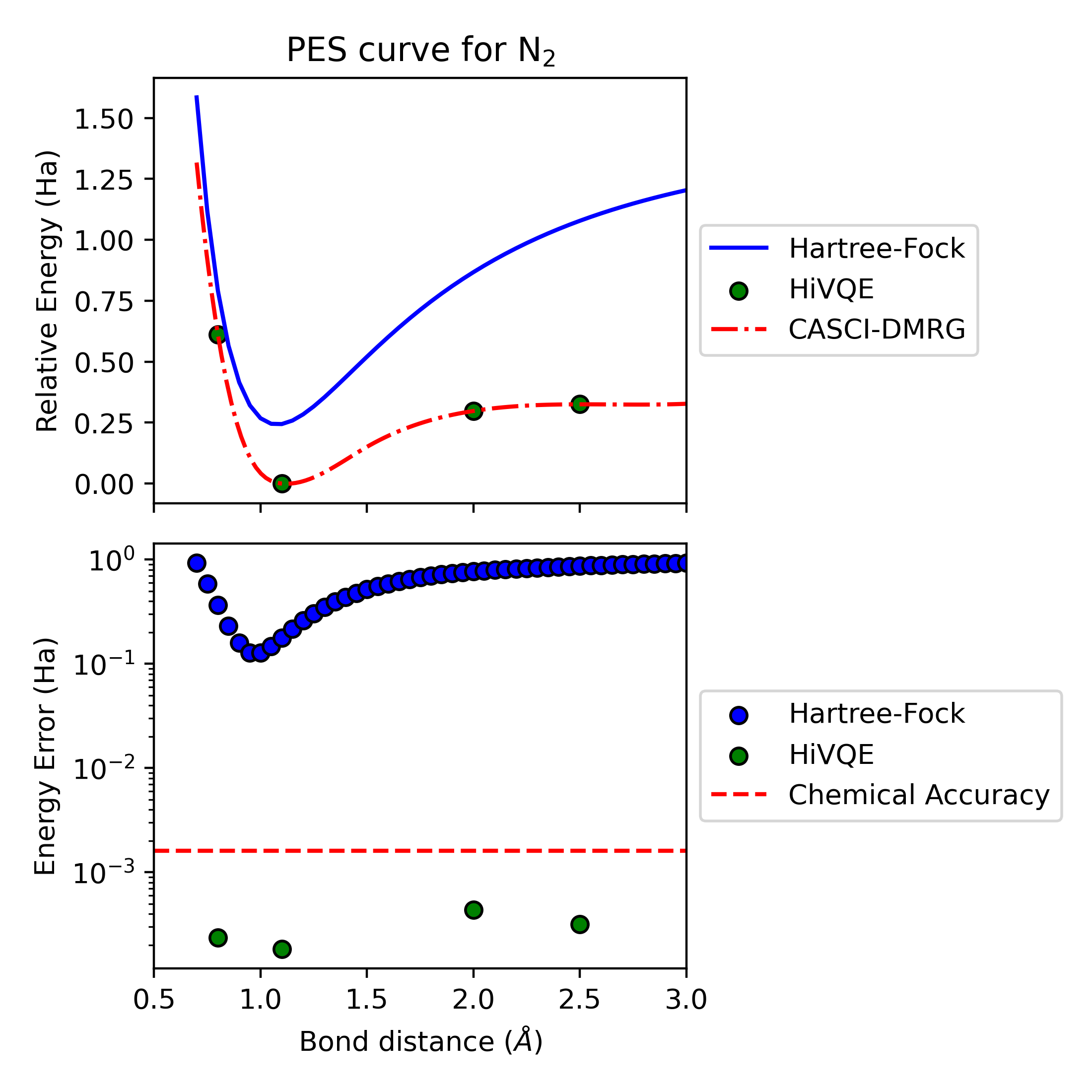}
    \end{adjustbox}
    \caption{The Potential Energy Surface (PES) and energy error analysis for the 
\ce{N2} with 40 qubits or (20o,14e). The top panel compares the relative energy computed using Hartree-Fock (blue line), HI-VQE (green circles), and CASCI-DMRG (red dashed line) as a reference with the corresponding geometries for each energy point at the top. The bottom panel presents the energy error relative to CASCI-DMRG on a logarithmic scale.}
    \label{fig:7}
\end{figure}
Figure \ref{fig:6} presents the potential energy surface (PES) of \ce{Li2S} system, illustrating the relative energy variations as a function of bond distance. The top panel compares three computational methods: Hartree-Fock (HF), HI-VQE, and CASCI. The Hartree-Fock PES deviates by more than hundreds of millihartrees from the CASCI reference (red dashed line), particularly at longer bond distances, indicating the limitations of mean-field approximations with a single HF reference state. In contrast, the HI-VQE results (green circles) closely follow the CASCI curve, demonstrating improved accuracy in capturing electron correlation effects.
The bottom panel shows the energy error relative to CASCI on a logarithmic scale. The Hartree-Fock errors remain consistently above the chemical accuracy threshold. Conversely, the HI-VQE errors stay well below the chemical accuracy threshold across all bond distances, confirming that HI-VQE achieves high precision and effectively reduces errors in quantum calculations of the
\ce{Li2S} dissociation process. Overall, the results demonstrate that HI-VQE significantly successfully captures the correlation energies, making it a promising approach for quantum simulations of complex chemical systems.

\subsubsection{Nitrogen}

The dissociation of the nitrogen molecule, \ce{N2}, presents a well-known challenge for quantum chemistry methods due to the interplay of dynamic and static electron correlation effects. As \ce{N2} transitions from a strongly bound state to dissociated atoms, the increasing multireference character necessitates advanced approaches such as multireference CI or coupled-cluster methods. While Hartree-Fock method fails to provide a qualitatively correct dissociation curve, post-HF correlation methods offer deeper insights into the nature of the bond-breaking process. Studying this system allows for a rigorous assessment of correlation energy treatments and serves as a benchmark for developing high-accuracy ab initio methodologies. As one of possible solution to this challenging problem, we performed HI-VQE calculation of \ce{N2} bond dissociation with 40 qubits, (20o,14e) (Figure \ref{fig:7}).

For systems with more than 40 qubits, solving them without a powerful high-performance computing (HPC) cluster is infeasible due to the immense memory required to store and diagonalize a full matrix with 6,009,350,400 states for (20o,14e). As a result, reference data for \ce{N2} using CASCI could not be obtained. To verify the accuracy of HI-VQE calculations, we instead computed the energies using CASCI-DMRG. The incremental nature of the HI-VQE method enables the identification of highly contributing states from the entire space while significantly reducing memory and classical computational power requirements.

Interestingly, we observed that the absence of the correlation energy in \ce{N2} dissociation can alter the conclusion of the chemical reaction, as shown in Figure \ref{fig:7}. This figure compares the energy difference between the equilibrium structure at 1.1 \AA{}  and 2.5 \AA, using the PES curves from HF and HI-VQE (or CASCI-DMRG). In the case of the HF study, bond dissociation occurs with nearly 1 Ha, while HI-VQE suggests that dissociation should be possible with only 0.25 Ha.

\subsubsection{Contracted subspace of HI-VQE method}

\begin{figure}[H]
\begin{adjustbox}{center}
    \includegraphics[width=0.6\textwidth,]{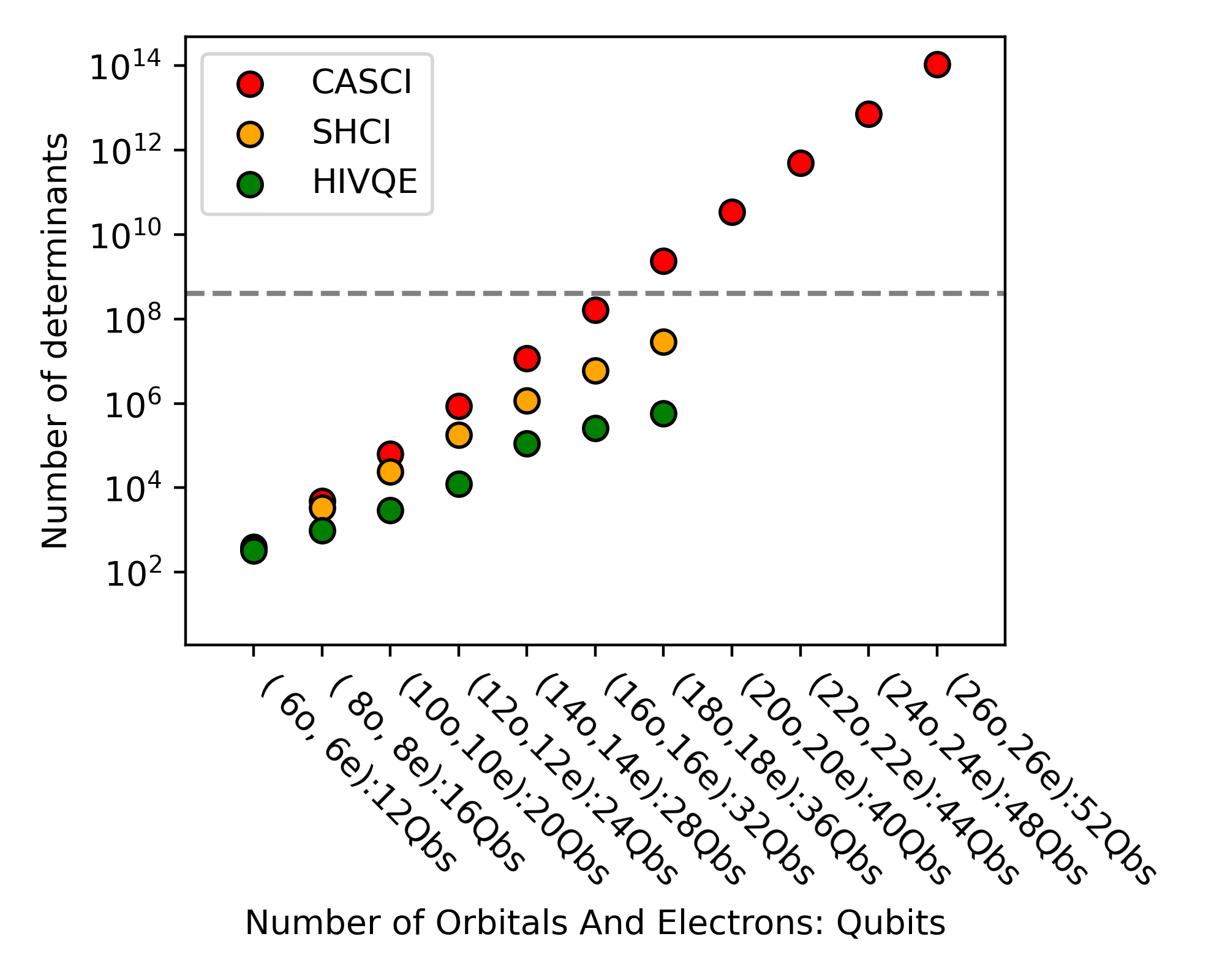}
    \end{adjustbox}
    \caption{Number of determinants vs the number of qubits obtained from three approaches (HI-VQE, SHCI and CASCI)}
    \label{fig:8}
\end{figure}

We conducted benchmark studies on several interesting chemical systems using the HI-VQE method. A key advantage of configuration interaction-based calculations is achieving the desired accuracy with fewer determinants and states, which grow exponentially in the ideal case. This is especially important when estimating physical and chemical properties from multireference wavefunctions, as it reduces the need to compute reduced density matrices in higher dimensions. Figure \ref{fig:8} illustrates the scaling of the number of determinants for CASCI (or FCI), SHCI, and HI-VQE methods. HI-VQE can identify subspaces with three orders of magnitude fewer states and two orders of magnitude fewer states than SHCI, all while achieving chemical accuracy.

\section{Conclusion}

In this work, we propose HI-VQE as a promising hybrid quantum-classical approach for addressing the multireference quantum chemistry problem. By effectively harnessing the strengths of both NISQ hardware and classical computation, HI-VQE delivers accurate and computationally efficient estimations of ground-state wavefunctions. This method not only mitigates quantum noise but also enables the generation of compact, yet chemically precise wavefunctions. The HI-VQE algorithm has been tested and validated across systems ranging from 16 to 40 qubits, including \ce{NH3}, \ce{Li2S} and \ce{N2}. The results show that HI-VQE can efficiently explore vast configuration spaces. As a result, HI-VQE has the potential to significantly accelerate advancements in quantum chemistry simulations, driving progress in materials discovery and chemical studies, while opening new avenues for exploring strongly correlated electronic systems and discovering novel materials.

\bibliography{references}

\end{document}